\documentclass[reprint,aps,prb,amsmath,amssymb,longbibliography,floatfix,superscriptaddress]{revtex4-2}

\usepackage{graphicx}
\usepackage{bm}
\usepackage[hidelinks, colorlinks=true, linkcolor=blue, citecolor=blue, urlcolor=blue]{hyperref}
\usepackage{orcidlink}
\usepackage{url}
\usepackage{comment}

\usepackage[utf8]{inputenc}
\usepackage[T1]{fontenc}
\usepackage{mathptmx}
\usepackage{etoolbox}

\makeatletter
\def\@email#1#2{%
	\endgroup
	\patchcmd{\titleblock@produce}
	{\frontmatter@RRAPformat}
	{\frontmatter@RRAPformat{\produce@RRAP{*#1\href{mailto:#2}{#2}}}\frontmatter@RRAPformat}
	{}{}
}%
\makeatother

\begin{document}

	\title{Fluctuation-Driven Enhancement of Spin-Orbit Torque near the Curie Temperature of Ultrathin Ferromagnets}

	\author{Mateusz Szurek\,\orcidlink{0009-0001-1032-1073}}
	\email{mateusz.szurek@emory.edu}
	\author{Sergei Ivanov\,\orcidlink{0000-0001-9594-6381}}
	\author{Sergei Urazhdin\,\orcidlink{0000-0002-6972-0300}}
	\affiliation{Department of Physics, Emory University, Atlanta, Georgia 30322, USA}

	\date{\today}


	\begin{abstract}
		We investigate how magnetic fluctuations influence spin-orbit torque in ultrathin-film magnetic heterostructures whose Curie temperature $T_C$ is suppressed by confinement. Above $T_C$, the damping-like contribution to spin-orbit field is significantly enhanced while the field-like contribution is suppressed, with the two contributions exhibiting opposite field dependencies. We show that these behaviors are consistent with fluctuation-driven mixing between the longitudinal and transverse interfacial spin conductances, which enhances absorption of transversely polarized spin current by the ferromagnet. This mechanism can be activated below $T_C$ by engineering the microscopic magnetic state and by harnessing spin current-generated short-wavelength magnons, suggesting a spintronic analog of heat-assisted magnetic recording.
	\end{abstract}

	\maketitle

	%===============================================================================
	\section{Introduction}
	%===============================================================================

	Spin-orbit torques (SOTs) provide a versatile mechanism for electronic control of nanomagnetic devices such as magnetic memory and logic, motivating the search for materials enabling efficient charge to spin conversion~\cite{Ramaswamy2018}. SOTs are usually generated by some combination of the spin Hall effect (SHE) in materials with strong spin-orbit coupling (SOC) and the Rashba-Edelstein effect (REE) at interfaces~\cite{Manchon2019}. Very large spin Hall angles have been reported for topological insulators~\cite{Wang2017}, but their power efficiency is limited by high resistivities. The smallest electric energy required to generate a unit of spin angular momentum is presently achieved with SHE sources based on alloys of heavy metals (HMs) such as Pt, Ta, and W, which still fall short of application requirements~\cite{Zhu2021}.

	Several recent developments hold promise for improvements in SOT efficiency. It was recently recognized that orbital counterparts of the SHE and REE can generate orbital torques supplementing the SOT~\cite{Jo2018,f8l4-rgnr}. Magnetic materials themselves can also serve as a source of spin current via the magnetic SHE and SOC-driven spin dynamics at magnetic interfaces~\cite{Manchon2008,Amin2016Phen}. These mechanisms offer a promising route for engineering hybrid spin-orbitronic heterostructures driven by a combination of bulk and interfacial sources of orbital and spin currents, although practical devices exploiting this hybrid approach have yet to be demonstrated.

	A complementary avenue for enhancing SOT is to engineer the magnetic state of the system itself. Since the effects of SOTs on magnetic systems scale inversely with magnetization, SOT efficiencies are expected to be enhanced in nearly compensated ferrimagnets (FIM)~\cite{Finley2016,Mishra2017} and antiferromagnets (AFMs)~\cite{Urazhdin2007,Zhou2019,Bose2025}. However, the SOT effects on magnetization dynamics are scaled by the magnetic relaxation rates, which are proportional to the characteristic dynamical frequencies and are significantly higher in nearly compensated FIMs and AFMs than in FMs. An additional constraint on SOT efficiency in AFMs is imposed by energy conservation in the magnon generation~\cite{Mitrofanov2020}. In FMs, the energies of long-wavelength magnons that dominate switching dynamics are at least two orders of magnitude smaller than the characteristic energies of electrons, and the effect of energy-conservation constraint on magnon generation is negligible. In contrast, in AFMs the two energy scales are comparable, resulting in the dominant role of energy conservation, rather than spin transfer, in spin current-driven magnetization dynamics~\cite{Mitrofanov2021}.
	  
	In this work, we address the role of the magnetic state of FM on SOT by studying the effects of magnetic fluctuations on SOTs near the Curie temperature $T_C$ of ultrathin ferromagnets (FMs). Traditional second-harmonic Hall voltage (HHV) analysis assumes a saturated magnetic state and a uniform anisotropy field $H_a$~\cite{Hayashi2014,Zhu2021,Lin_Zhu_2024}. These assumptions break down in ultrathin FMs near $T_C$ due to short-range magnetic fluctuations and local variations of anisotropy due to its scaling by the film roughness, which in sputtered ultrathin transition-metal (TM) FMs can be comparable to film thickness. We bypass the requirement of magnetic saturation by reformulating the response in terms of generalized magnetoelectronic susceptibilities (GMS), which allows the direct extraction of effective SOFs even in the presence of strong fluctuations. Applying this method to ultrathin CoFeB/Pt and CoFeB/Ta bilayers, we uncover an anomalous enhancement of the damping-like (DL) and suppression of field-like (FL) SOT efficiencies near $T_C$. Both effects are consistent with fluctuation-driven mixing between the longitudinal and transverse channels of the interfacial spin conductance. The demonstrated effect provides an approach for SOT-driven devices harnessing Joule heating and short-wavelength magnons generated directly by spin current~\cite{Demidov2011}, for improved-efficiency operation analogous to heat-assisted recording.

    The paper is organized as follows. In the next two sections, we outline the mechanism of mixing between longitudinal and transverse spin conductance by microscopic magnetic inhomogeneity that results in enhancement of DL-SOT and suppression of FL-SOT. In Section~\ref{sec:GMS}, we introduce the generalized magnetoelectronic susceptibility (GMS) method of SOT characterization in inhomogeneous magnetization states used in this work, and describe measurements for a control sample validating the method. In Section~\ref{sec:Characterization}, we provide the details on fabrication and characterization of the studied ultrathin FM films and nanopatterned samples. The SOT measurements are described in Section~\ref{sec:Results} and discussed in Section~\ref{sec:Discussion}. Finally, Section~\ref{sec:Summary} summarizes our findings.
	
    \section{Effect of fluctuations on the spin-orbit torque}
	\label{sec:Theory}

	To analyze the effects of magnetization state on SOTs, we consider SOT expressed in terms of effective SOF $\bm{H}_{\mathrm{SOT}}=\bm{H}_{\mathrm{FL}}+\bm{H}_{\mathrm{DL}}$, where $\bm{H}_{\mathrm{FL}}$ and $\bm{H}_{\mathrm{DL}}$ are the effective field-like and damping-like SOFs, respectively~\cite{Manchon2019},
	\begin{equation}\label{eq:torque}
		\bm{\tau}=-\gamma\mu_0 M_s\,\bm{m}\times\bm{H}_{\mathrm{SOT}}.
	\end{equation}
	Here, $\bm{m}$ is a unit vector along the local magnetization and $M_s$ is its magnitude. Equation~(\ref{eq:torque}) shows that the SOT is dependent on the orientation of the uniform magnetization, becoming maximized when the effective fields are orthogonal to $\bm{m}$. This geometric optimum is not directly useful for SOT-driven magnetization switching or precession usually driven by SOF nearly collinear with $\bm{m}$.
    
    To analyze the effects of magnetic inhomogeneity, we consider the expression for SOFs dominated by the SHE,
	\begin{equation}\label{eq:effective_fields}
		H_{\mathrm{DL}}+i\,H_{\mathrm{FL}}
		=\frac{\hbar\,\xi_{\mathrm{SH}}\,j_c}{2e\mu_0 M_s t_{\mathrm{FM}}},
	\end{equation}
	where $t_{\mathrm{FM}}$ is the FM thickness, $j_c$ is the charge current density in the SHE layer, and $\xi_{\mathrm{SH}}=T_{\mathrm{int}}\theta_{\mathrm{SH}}$ is a complex spin Hall efficiency determined by the spin Hall angle $\theta_{\mathrm{SH}}$ and the complex interface transparency is given by~\cite{Zhu2021}
	\begin{equation}\label{eq:Tint}
		T_{\mathrm{int}}=\frac{2G_{\mathrm{mix}}\tanh(t_N/2\lambda_s)}{2G_{\mathrm{mix}}\coth(t_N/\lambda_s)+G_N}.
	\end{equation}
	Here $t_N$ is the thickness of the SHE layer, $\lambda_s$ is its spin-diffusion length, $G_N\approx\lambda_s/\rho_N$ is the effective bulk spin conductance, $\rho_N$ is the resistivity, and $G_{\mathrm{mix}}=G_r+iG_i$ is the complex interface spin-mixing conductance. Typically $G_{i,r}\ll G_N$, then the SOFs scale approximately linearly with $G_{i,r}$.

	The real and imaginary parts of $G_{\mathrm{mix}}$ are the off-diagonal elements of the spin-conductance tensor in a reference frame with $\hat{z}\parallel\bm{m}$,
	\begin{equation}\label{eq:cond_tensor}
		\bm{G}_0=\begin{pmatrix}
			G_r & -G_i & 0\\
			G_i &  G_r & 0\\
			0   &   0  & G_P
		\end{pmatrix},
	\end{equation}
	where $G_P$ is the longitudinal interface spin conductance. For clean interfaces between similar materials such as TMs, $G_P$ is close to the Sharvin conductance, and $G_r,\,|G_i|<G_P$.

	The high current densities required for SOT-driven magnetization switching or precession inevitably produce Joule heating, as well as spin current-driven magnetic fluctuations not limited to quasi-uniform precession~\cite{Demidov2011,Demidov2017}. We consider the limiting case in which the characteristic length scale of these fluctuations is shorter than the spin-diffusion length, so that the relevant spin transport properties of the interface are determined by the average of $\bm{G}_0$ over the microscopic magnetic configurations. To establish the effect on the transverse spin conductance, we transform the tensor in Eq.~(\ref{eq:cond_tensor}) into the frame defined by the \emph{average} direction $\langle\bm{m}\rangle$ of the fluctuating magnetization. The instantaneous orientation is described by the position- and time-dependent spherical angles $(\theta_f,\varphi_f)$ relative to that frame. Assuming that $\varphi_f$ is uniformly distributed, and averaging over the distribution of $\theta_f$, we obtain
	\begin{equation}\label{eq:cond_tensor_av}
        \langle\bm{G}\rangle=
        \begin{pmatrix}
			G_r+\tfrac{D}{2}\langle\sin^2\theta_f\rangle & -G_i\langle\cos\theta_f\rangle & 0\\
			G_i\langle\cos\theta_f\rangle & G_r+\tfrac{D}{2}\langle\sin^2\theta_f\rangle & 0\\
			0 & 0 & G_r+D\langle\cos^2\theta_f\rangle
		\end{pmatrix},
    \end{equation}
	where $D=G_P-G_r$.

	Equation~(\ref{eq:cond_tensor_av}) describes fluctuation-induced mixing between the longitudinal and transverse channels of the interfacial spin conductance. The imaginary mixing component is reduced by $\langle\cos\theta_f\rangle\le 1$, while the real component $G'_r=G_r+\tfrac{D}{2}\langle\sin^2\theta_f\rangle$  is increased if $G_P>G_r$. Physically, the spin component transverse to $\langle\bm{m}\rangle$ contains a component collinear with the local magnetization, which is efficiently injected into FM due to the large longitudinal conductance. As a result, the DL torque is expected to be enhanced. In contrast, the FL torque, which originates from the coherent precession of injected spins about the local exchange field, is suppressed because that precession is dephased by the fluctuations.

	Enhancement of the DL-SOT is desirable since it dominates current-driven magnetization switching and spin-wave generation in magnonic applications~\cite{Manchon2019,Demidov2020}. The proposed longitudinal-transverse mixing mechanism is not restricted to thermal fluctuations: it can also be activated below $T_C$ by engineering the static microscopic magnetic state, for example through frustrated exchange coupling across a ferromagnet/antiferromagnet (AFM) interface~\cite{Chen2020}. This may have contributed to the DL-SOT enhancement reported upon insertion of an ultrathin AFM NiO spacer at the magnetic interface~\cite{Zhu2021b}, and to the enhanced SHE observed when a near-critical magnetic material was used as the source~\cite{PhysRevLett.120.097203}, which were originally attributed entirely to magnon-mediated torques. Below, we describe the study of SOTs near the Curie temperature of the magnetic layer itself, where no distinction between magnon-mediated and conventional SOT can be made, allowing us to isolate the essential geometric role of an inhomogeneous magnetic state as opposed to dynamic magnon transport.

	%===============================================================================
	\section{Magnetoelectronic susceptibility method for SOF characterization}
	\label{sec:GMS}
	%===============================================================================

	The standard HHV analysis of SOTs assumes a saturated state with uniform anisotropy field $H_a$~\cite{Zhu2021,Lin_Zhu_2024,Hayashi2014}. In this framework, the in-plane angular dependence of the rms second-harmonic voltage takes the form
	\begin{equation}\label{eq:harmonic_1}
		V_{2\omega}(\varphi)=
		\frac{I_0}{\sqrt{2}}
		\left[
		\frac{R_{\mathrm{AHE}}\,H_{\mathrm{DL}}}{H+H_a}
		+
		2R_{\mathrm{PHE}}\,\frac{H_{\mathrm{FL}}+H_{\mathrm{Oe}}}{H}\cos 2\varphi
		\right]\cos\varphi,
	\end{equation}
	where $I_0$ is the rms amplitude of ac current, $R_{\mathrm{AHE}}$ and $R_{\mathrm{PHE}}$ are the amplitudes of the anomalous and planar Hall effects (AHE and PHE), $H_{\mathrm{DL}}$ and $H_{\mathrm{FL}}$ are the rms SOF amplitudes, and $H_{\mathrm{Oe}}$ is the Oersted field.	Planar and anomalous Nernst contributions to $V_{2\omega}$ are negligible for the ultrathin magnetic films considered here, where normal to the film thermal gradients are small since thermal conductance is dominated by the Kapitza resistance~\cite{PhysRevApplied.14.024024}.

	\begin{figure}
		\includegraphics[width=\columnwidth]{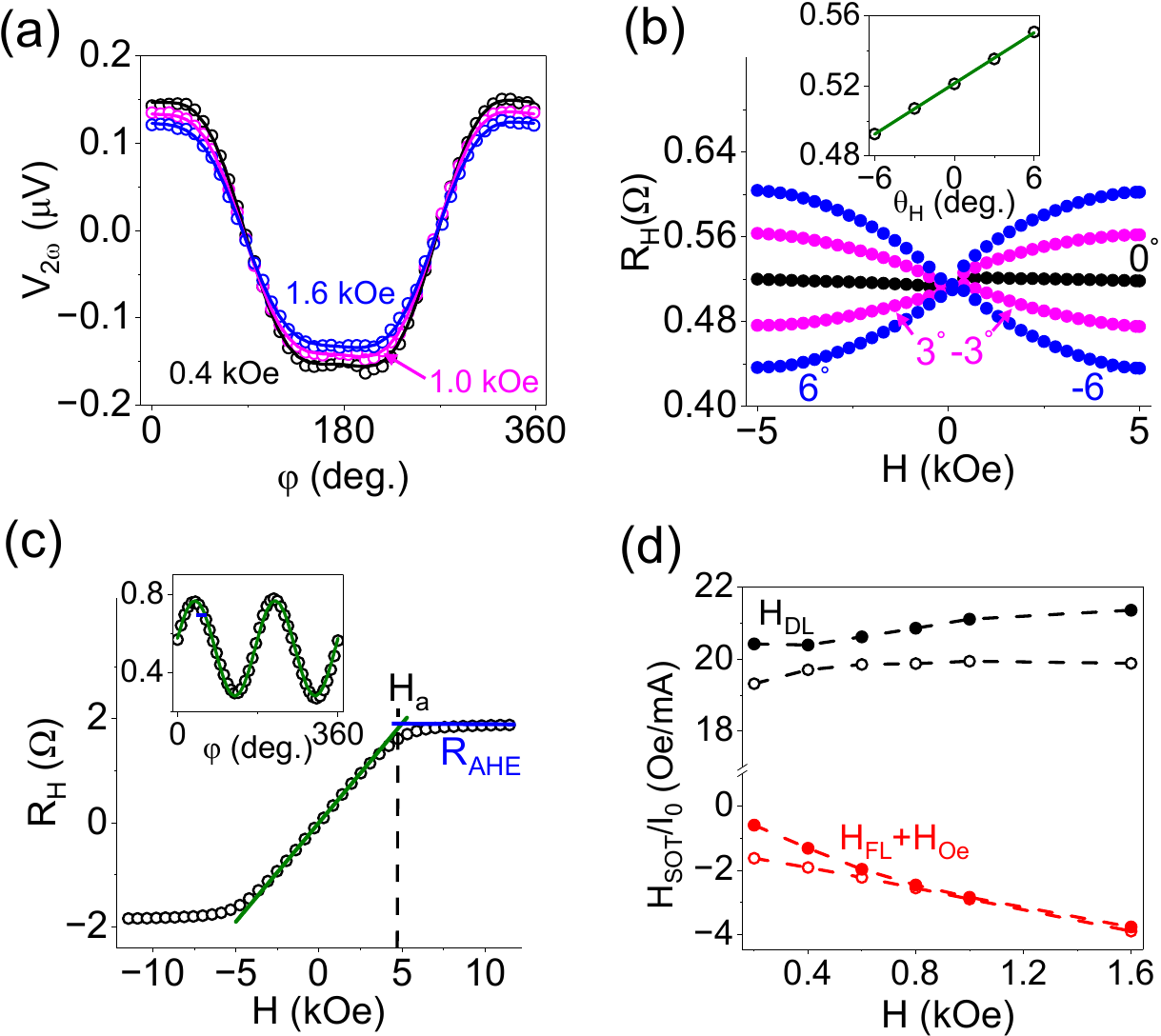}
		\caption{\label{fig:1} 
			Cross-validation of the GMS and standard HHV methods on a CoFeB(1.3)/Pt(2.5) control sample at $T=300$~K. (a) Second-harmonic voltage $V_{2\omega}$ vs in-plane field angle $\varphi$ at the labeled in-plane field magnitudes, performed using $I_0=0.2$~mA. (b) $R_H(H)$ for $\theta_H=\pm 3^\circ,\pm 6^\circ$. Inset: $R_H(\theta_H)$ at $H=1$~kOe. (c) AHE loop used to extract $R_{AHE}$ and $H_a$; the red line is a linear fit near zero field, blue line is the saturation value. Inset: PHE used to extract $R_{PHE}$ via $R_H(\varphi)=R_{PHE}\sin(2\varphi)$. (d) Normalized effective SOT fields determined by the standard HHV (open symbols) and GMS method (solid symbols) vs in-plane field magnitude.}
	\end{figure}

	Equation~(\ref{eq:harmonic_1}) is not directly applicable near $T_C$, where thermal magnetic fluctuations result in a microscopically inhomogeneous state that cannot be described by a unique value of $\varphi$. Additionally, the scaling of interfacial anisotropy by the inhomogeneous film thickness  in ultrathin magnetic films results in non-negligible local variations of $H_a$. To overcome these limitations, we introduce a self-consistent generalized magnetic susceptibility (GMS) approach, which replaces the field- and anisotropy-dependent prefactors in Eq.~(\ref{eq:harmonic_1}) with directly measured Hall susceptibilities. Throughout, $\theta$ ($\theta_H$) denotes the angle of $\bm{m}$ ($\bm{H}$) with respect to the film plane, and $\varphi$ is the in-plane field angle measured from the current direction. The in-plane anisotropy of FM (amorphous CoFeB in our study) is neglected. The second-harmonic voltage can then be written as
	\begin{equation}\label{eq:harmonic_2}
		V_{2\omega}(\varphi)=
		\frac{I_0}{\sqrt{2}}\left[
		H_{\mathrm{DL}}\,\frac{dR_H}{dH_{\mathrm{OP}}}
		+
		(H_{\mathrm{FL}}+H_{Oe})\,\frac{dR_H}{dH_{\mathrm{IP}}}\cos 2\varphi
		\right]\cos\varphi,
	\end{equation}
	where $dR_H/dH_{\mathrm{OP}}$ and $dR_H/dH_{\mathrm{IP}}$ are the susceptibilities of the Hall resistance with respect to out-of-plane and in-plane transverse perturbations of the magnetic field when the latter is aligned with the current ($\varphi=0$). This formulation requires no \emph{a priori} knowledge of the magnetic state, drawing only on the directly measured magnetoelectronic response.

	To benchmark the GMS approach against the standard HHV analysis, we performed a comparative study of a control sample of CoFeB(1.3)/Pt(2.5) at room temperature (RT) $T=300$~K, fabricated as described in the next section. Numbers in parentheses are thicknesses in nanometers. The chosen CoFeB thickness is sufficiently large so that its $T_C$ is well above RT, while remaining in the ultrathin regime such that transport properties are not significantly different from the other studied films with lower $T_C$. Consequently,  the relative thickness variations of the film remain non-negligible.

	The dependences $V_{2\omega}(\varphi)$ for three representative magnitudes of in-plane field are shown in Fig.~\ref{fig:1}(a). To characterize GMS, we determined $R_H(H)$ at small out-of-plane and in-plane tilts, as illustrated for out-of-plane tilt in Fig.~\ref{fig:1}(b)). As expected, the dependence is linear at small tilt angles (inset). The susceptibilities are then obtained as
	\begin{equation}\label{eq:chi_finite_diff}
		\frac{dR_H}{dH_{\mathrm{OP}}}\approx\frac{R_H(\theta)-R_H(-\theta)}{2H\theta},\quad
		\frac{dR_H}{dH_{\mathrm{IP}}}\approx\frac{R_H(\varphi)-R_H(-\varphi)}{2H\varphi}.
	\end{equation}
	For the comparison, we independently measured $H_a$ and $R_{\mathrm{AHE}}$ from AHE loops (Fig.~\ref{fig:1}(c)), $R_{\mathrm{PHE}}$ from in-plane angular sweeps (inset in Fig.~\ref{fig:1}(c)). The normalized rms Oersted field computed using Amp\`ere's law,  $H_{\mathrm{Oe}}/I_0\sim2\pi\mu_0/w=1.2$~Oe/mA is significantly smaller than the typical SOFs reported below, and is neglected in our subsequent analysis.

	Figure~\ref{fig:1}(d) shows the effective fields extracted from the GMS method (Eq.~(\ref{eq:harmonic_2})) and from the standard HHV analysis (Eq.~(\ref{eq:harmonic_1})). At fields $H>400$~Oe, the SOFs calculated by the two methods agree to within $10\%$. The increasing discrepancy at small fields can be attributed to the effects of random in-plane anisotropy induced in CoFeB by its interface with Pt. The GMS method is agnostic to these effects, since the in-plane anisotropy is captured by the response to in-plane field tilt. Meanwhile, the standard HHV analysis in the form Eq.~(\ref{eq:harmonic_1}) does not account for these effects. Notably, both the standard HHV and GMS methods show a decrease of $H_{FL}$ at small $H$, which is consistent with the effects of nanoscale magnetic inhomogeneity discussed below.

	%===============================================================================
	\section{Sample fabrication and magnetic characterization}
	\label{sec:Characterization}
	%===============================================================================

To test the proposed effects of fluctuations on SOTs, we studied ultrathin-film heterostructures based on the prototypical FM/HM bilayers CoFeB/Pt and CoFeB/Ta~\cite{Pai2012,PhysRevApplied.14.024024}. The films were deposited on sapphire by ultrahigh vacuum sputtering, and patterned  for magnetoelectronic measurements into $5\,\mu m$-wide Hall bars. The thickness $t_{FM}$ of amorphous CoFeB=Co$_{40}$Fe$_{40}$B$_{20}$ was adjusted so that its Curie temperature was reduced from the bulk value $T_C=900$~K to $150-250$~K due to the quantum confinement~\cite{Zhang2001,Vaz2008}, in the temperature range accessible in the measurements described below. 

Proximity magnetism in Pt enhances magnetic ordering in FM/Pt bilayers~\cite{Lim2013}. In contrast, FM/Ta bilayers exhibit a magnetically dead layer~\cite{Cecot2017}. Accordingly, the desired range of $T_C$ was achieved with $t_{FM}=0.65$~nm for CoFeB/Pt, smaller than $t_{FM}=1.1$~nm for CoFeB/Ta. The film surface roughness of about $0.3$~nm RMS characterized by atomic force microscopy was similar to that of the substrate (see Supplemental Material \cite{supplemental}). Nevertheless, ultrathin magnetic layers are likely not fully continuous, resulting in finite-size effects and local variations of magnetic anisotropy. These effects were included in our analysis as described below. Despite substantial differences in film thicknesses and the opposite SHE signs of Ta and Pt~\cite{Liu2012, Pai2012}, the studied structures exhibited similar current-driven behaviors close to the Curie point, confirming that the observed effects are a general consequence of fluctuations close to $T_C$. 

	\begin{figure}[t]
		\includegraphics[width=\columnwidth]{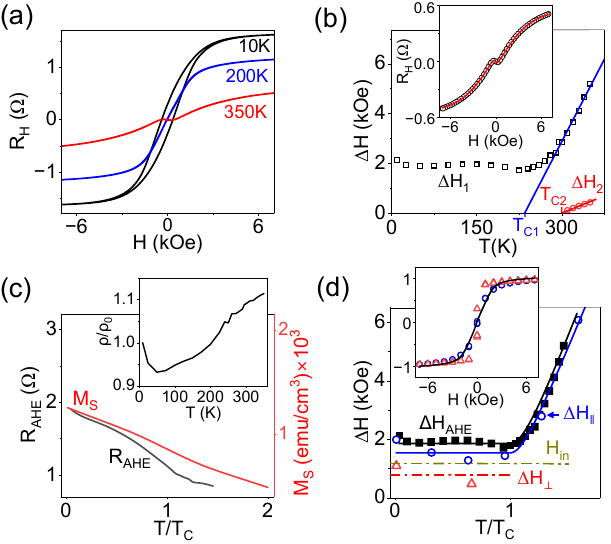}
		\caption{\label{fig:2}
			Magnetic characterization of the CoFeB(0.65)/Pt(2.5) Hall bar.
			(a) Hall resistance vs out-of-plane field at the labeled temperatures.
			(b) Broadening $\Delta H$ of $R_H(H)$ vs $T$ extracted from two-sigmoid fits, with extrapolated critical temperatures $T_{C1}$ and $T_{C2}$ obtained from the Curie--Weiss law for the respective sigmoids. Inset: example two-sigmoid fit at $T=350$~K.
			(c) Saturation magnetization $M_s$ from SQUID magnetometry (red) and AHE amplitude $R_{\mathrm{AHE}}$ (black) vs $T/T_C$. Inset: normalized resistivity $\rho/\rho_0$ vs $T$; the upturn at low $T$ is attributed to the Kondo effect.
			(d) Broadening $\Delta H$ of the AHE and $M(H)$ curves obtained from the sigmoid fits for in-plane and out-of-plane fields. Inset: $R_H(H)$ and $M(H)$ at $T=0.7\,T_C$ for in-plane (circles) and out-of-plane (triangles) fields. Magnetoelectronic data are from the CoFeB(0.65)/Pt(2.5) bilayer on sapphire patterned into a $5\,\mu$m Hall bar; SQUID data are from an extended Ti(2)/Pt(2)/CoFeB(0.32)/AlO$_x$(4) film on polyimide.}
	\end{figure}

	Figure~\ref{fig:2}(a) shows representative AHE curves for a CoFeB(0.65)/Pt(2.5) Hall bar. At cryogenic $T$ the curves display a small hysteresis that closes as $|R_H|$ saturates for $|H|>2$~kOe, consistent with the ordered state of an ultrathin film whose stability is limited by finite-size effects, thermally activated domain-wall motion, and a distribution of local anisotropies~\cite{Zhang2001}. As $T$ increases, the AHE amplitude decreases and the curves broaden, indicating a transition to the paramagnetic regime. Above $T\!\approx\!300$~K an additional nonlinear feature emerges at small $H$. Similar features in other ultrathin FM/HM heterostructures have been attributed to ``Rashba magnetism'', i.e., incipient magnetic ordering driven by the competition between Rashba SOC and exchange interaction~\cite{Golod2013,PhysRevApplied2024,Barnes2014,Kim2016}. The full $R_H(H)$ curve is well described by a sum of two sigmoids (inset of Fig.~\ref{fig:2}(b)),
	\begin{equation}\label{eq:two_sigmoid}
		R_H(H)=R_{\mathrm{AHE}}\tanh\!\left(\frac{H}{\Delta H_1}\right)
		+R'_{\mathrm{AHE}}\tanh\!\left(\frac{H}{\Delta H_2}\right),
	\end{equation}
	whose widths $\Delta H_{1,2}$ are inverse magnetoelectronic susceptibilities normalized by the corresponding Hall amplitudes. As shown in Fig.~\ref{fig:2}(b), the widths follow the Curie--Weiss law, with extrapolated Curie temperatures $T_{C1}=250$~K for the conventional contribution that persists at low temperatures, and $T_{C2}=300$~K for the anomalous feature. For brevity we will refer to $T_{C1}$ simply as the Curie temperature, and use the label $T_{C}$ interchangeably with $T_{C1}$. The nearly constant sigmoid width below $T_{C1}$ rules out superparamagnetic blocking of discontinuous magnetic regions. Furthermore, the linear dependence $\Delta H_1(T)\propto(T-T_{C1})$ at $T>T_{C1}$ implies that the effective fluctuating moment $\mu\sim k_B T/\Delta H$ decreases with increasing $T$ above $T_C$. Consequently, at temperatures sufficiently above $T_{C1}$ the fluctuations satisfy the short-lengthscale assumption underlying Eq.~(\ref{eq:cond_tensor_av}). 

	As discussed in the previous section and illustrated by the rounded $R_H(H)$ curves in Fig.~\ref{fig:2}(a), ultrathin films near $T_C$ cannot be characterized by a well-defined anisotropy field, preventing analysis of SOFs using the standard HHV approach. The SOFs can be determined self-consistently from the GMS method introduced above, bypassing this issue. SOFs represent the fundamental SOT efficiencies scaled by the magnetization whose precise characterization is challenging due to exceedingly small magnetic moments of ultrathin magnetic films. In particular, our SQUID magnetometry measurements showed that the magnetic response of ultrathin films deposited on sapphire silicon substrates was dominated by artifacts from the trace amounts of magnetic impurities in the substrates and the sample holder. Thus, to estimate $M(H,T)$, we performed SQUID magnetometry on a Ti(2)/Pt(2)/CoFeB(0.32)/AlO$_x$(4) multilayer deposited on a $25\,\mu$m-thick polyimide film, which minimized the magnetic background from the substrate (see Supplemental Material \cite{supplemental}). A Ti(2)/Pt(2) buffer was used to ensure adhesion and magnetic continuity, and CoFeB thickness was reduced to compensate for the larger proximity magnetism effect at the Pt/FM interface. The film was cut into pieces sized to fit the SQUID sample space and glued with GE varnish into a stack of about 50 layers to amplify the magnetic signal relative to the background from the sample holder; an identically prepared bare polyimide stack was used for background subtraction. The Curie temperature of this stack estimated from sigmoid fits of $M(H,T)$ is $T_{C1}=150$~K.

	At cryogenic $T$ the saturation magnetization approaches $M_s=1.25\times 10^3\,\mathrm{emu/cm^3}$, close to the bulk CoFeB value, and decreases approximately linearly with increasing $T$ (Fig.~\ref{fig:2}(c)). Despite the differences in stack structure, the normalized AHE amplitude $R_{\mathrm{AHE}}(T/T_C)/R_{\mathrm{AHE}}(0)$ tracks $M_s(T/T_C)/M_s(0)$ to within $20\%$ over the full measured range. This is consistent with a ``dirty'' transport regime in which the AHE is dominated by the temperature-independent intrinsic and side-jump contributions~\cite{RevModPhys.82.1539}, so that $\sigma_{\mathrm{AHE}}\propto M_\perp(H,T)$. Combined with the small residual resistivity ratio of $\sim 1.1$ (inset of Fig.~\ref{fig:2}(c)), this implies $\rho_{\mathrm{AHE}}=\sigma_{\mathrm{AHE}}\rho^2\propto M_\perp(H,T)$. Consequently $R_H(H,T)$ provides a direct proxy for the out-of-plane magnetization component.

	To evaluate $M_\parallel(H,T)$ for in-plane fields based on the measurements of $R_H(H,T)$, the effect of the magnetic anisotropy and its inhomogeneity must be included. We adopt an empirical sigmoidal parametrization,
	\begin{equation}\label{eq:empiric_M}
		M_{\perp,\parallel}(H)=M_s(T)\tanh\!\bigl(H/\Delta H_{\perp,\parallel}\bigr),
	\end{equation}
	with widths
	\begin{equation}\label{eq:width}
		\Delta H_{\perp,\parallel}=\sqrt{a^2\,\Theta(T-T_C)(T-T_C)^2+(H_a\pm\Delta H_{\mathrm{in}})^2},
	\end{equation}
	where $H_a$ is the average anisotropy field, $\Delta H_{\mathrm{in}}$ accounts for additional broadening from finite-size effects and the inhomogeneous anisotropy distribution, $\Theta$ is the Heaviside step function, and $a$ is a fitting constant.

	This form captures the crossover between the Curie--Weiss law at $T>T_{C1}$ and the temperature-independent broadening at $T<T_{C1}$ (see an example of the AHE fit in Fig.~\ref{fig:2}(d)). Both SQUID and AHE data indicate that the anisotropy is nearly $T$-independent below $T_C$. The SQUID data for the films deposited on the kapton substrate follow Eq.~(\ref{eq:empiric_M}) with a different $H_a$, likely due to a small perpendicular anisotropy induced by the Ta/Pt buffer. Comparing in-plane and out-of-plane SQUID data below $T_{C1}$ yields $\Delta H_{\mathrm{in}}=0.4$~kOe. Using the same $\Delta H_{\mathrm{in}}$ for the micropatterned Hall bars, we obtain $M_\parallel(H)$ from the AHE data via Eq.~(\ref{eq:empiric_M}). Because of the uncertainty in $\Delta H_{\mathrm{in}}$, this estimate is semi-quantitative below $T_C$, but it becomes increasingly accurate above $T_C$ where $M(H)$ is dominated by fluctuations and the role of anisotropy is reduced.

	%===============================================================================
	\section{Results}
	\label{sec:Results}
	%===============================================================================

	\subsection{Harmonic measurements}

		\begin{figure}[t]
		\includegraphics[width=\columnwidth]{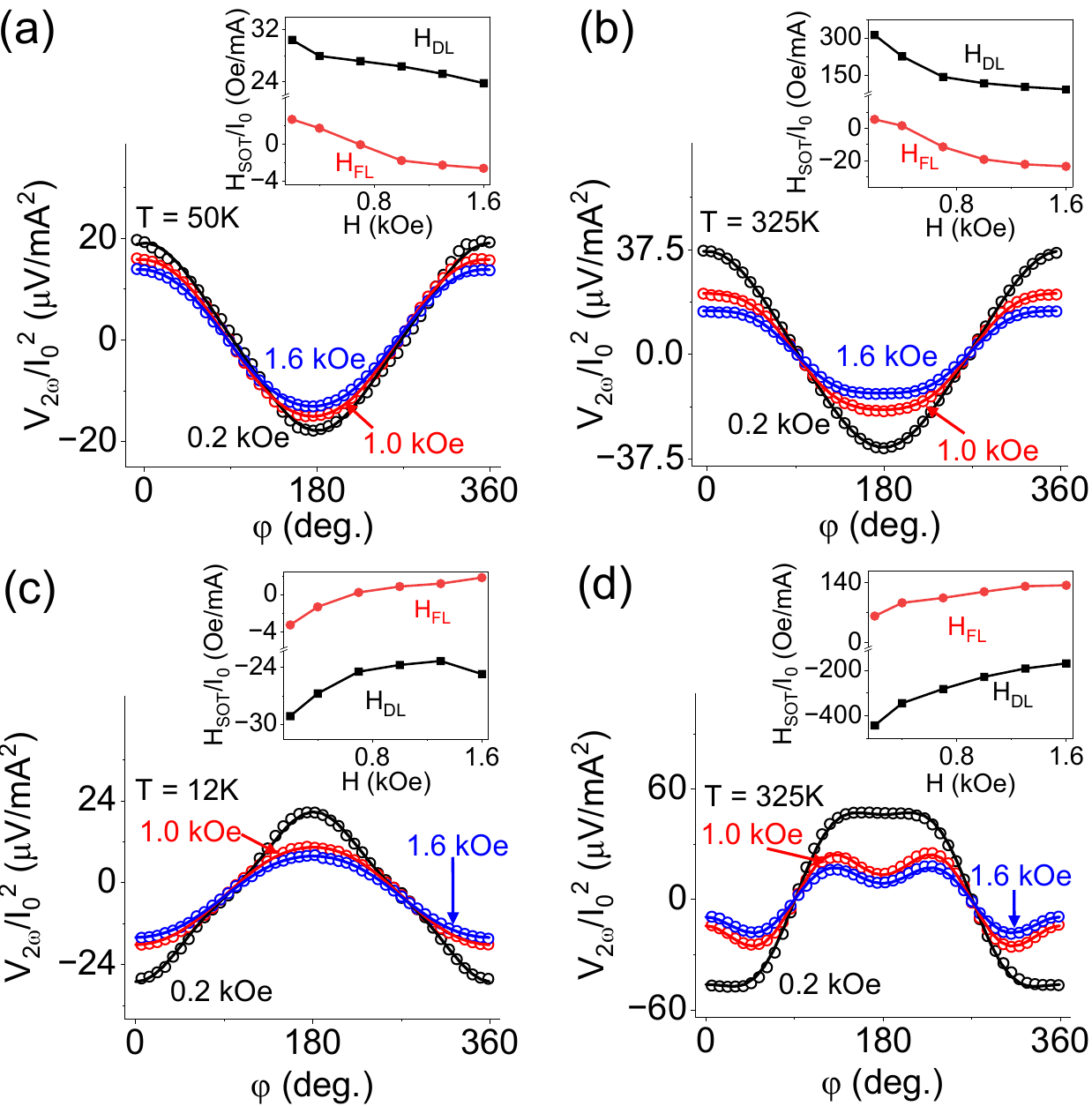}
		\caption{\label{fig:3}
			Normalized second-harmonic voltage $V_{2\omega}/I_0^2$ vs in-plane field angle $\varphi$ for CoFeB(0.65)/Pt(2.5) (a,b) and CoFeB(1.1)/Ta(4) (c,d) at the labeled field magnitudes and temperatures, with rms current amplitudes $I_0=0.2$~mA and $0.5$~mA, respectively. Curves are fits with Eq.~(\ref{eq:harmonic_2}). Insets: field dependence of the damping-like ($H_{\mathrm{DL}}$) and field-like ($H_{\mathrm{FL}}$) effective SOFs at the labeled $T$.}
	\end{figure}

	We characterized the SOTs in ultrathin films using the GMS method in the Hall geometry, by analyzing the dependence of the second-harmonic mixing voltage $V_{2\omega}$ on the in-plane field angle $\varphi$ with the current $I=\sqrt{2}I_0\cos(2\pi f t)$ at frequency $f=1.3$~kHz applied to the Hall bar. Representative dependences $V_{2\omega}(\varphi)$ are shown for the CoFeB(0.65)/Pt(2.5) sample in Fig.~\ref{fig:3}(a,b). In the ordered state at $T=50$~K (Fig.~\ref{fig:3}(a)), the angular dependence is nearly sinusoidal and only weakly dependent on $H$, with an amplitude that decreases monotonically with $H$ as expected from Eq.~(\ref{eq:harmonic_1}) for field-independent SOFs. Indeed, the effective SOFs extracted from these curves using the GMS method are almost constant at $H>0.4$~kOe. The FL-SOF is smaller than DL-SOF by an order of magnitude and at $H\gtrsim0.4$~kOe has the same sign as in the control sample with a thicker FM at room temperature (Fig.~\ref{fig:1}). Its magnitude decreases at small $H$, and the direction is reversed at $H=0.2$~kOe. Assuming that the physical mechanisms of SOTs do not qualitatively change with field, the sign reversal can be attributed to an additional competing  contribution with the opposite sign, which may originate from some combination of the interfacial Rashba-Edelstein effect~\cite{MihaiMiron2010}, Oersted field, orbital Hall contribution, and the effect of spin backflow. We note that in the sign convention used in this work the Oersted field has a positive sign, but its estimated normalized magnitude $H_{Oe}/I\approx 1.2$~Oe/mA is too small to fully account for the sign reversal. Furthermore, we observe a similar sign reversal at small $H$ at cryogenic temperatures in CoFeB/Ta (Fig.~\ref{fig:3}(c)), which cannot be explained by Oersted field.
		
	In contrast to the $T=50$~K results, at $T=325$~K above both the Curie temperature $T_{C1}=250$~K and $T_{C2}=300$~K, the angular dependence noticeably varies with the field magnitude. At small field, the dependence is almost sinusoidal, while at large fields the peaks of the curves are flattened, indicating an increased relative FL contribution. Indeed, the magnitude of FL-SOFs extracted from the fits using the GMS method Eq.~(\ref{eq:harmonic_2}) nearly vanishes at $H=0.2$~kOe but increases with $H$ and reaches about $1/3$ of $H_{DL}$ at $H=1.6$~kOe. Meanwhile, $H_{DL}$ decreases in this field range by a factor of $3$.
	
	The field-dependences of SOFs for the CoFeB(1.1)/Ta(4) sample, Fig.~\ref{fig:3}(c,d), are remarkably similar except for the flipped sign of both SOFs, consistent with the opposite signs of $\theta_{SH}$ in Pt and Ta.	In the ordered state at $T=12$~K, $H_{FL}$ is negligible at $H>0.2$~kOe while $H_{DL}$ is independent of $H$, with the magnitudes of both effective fields slightly increasing at $H=0.2$~kOe (Fig.~\ref{fig:3}(c)), exactly as in the CoFeB(0.65)/Pt(2.5) sample. At $T=325$~K above the Curie temperature $T_{C1}=300$~K, $H_{FL}$ is comparable to $H_{\mathrm{DL}}$ at $H=1.6$~kOe and decreases with decreasing $H$, extrapolating to a value close to $0$ at $H=0$, the same as for the CoFeB(0.65)/Pt(2.5). Meanwhile, $H_{DL}$ exhibits an opposite dependence, increasing by a factor of $4$ as $H$ is decreased from $1.6$~kOe to $0.2$~kOe, also similar to the Pt-based sample.

	\begin{figure}[t]
		\includegraphics[width=\columnwidth]{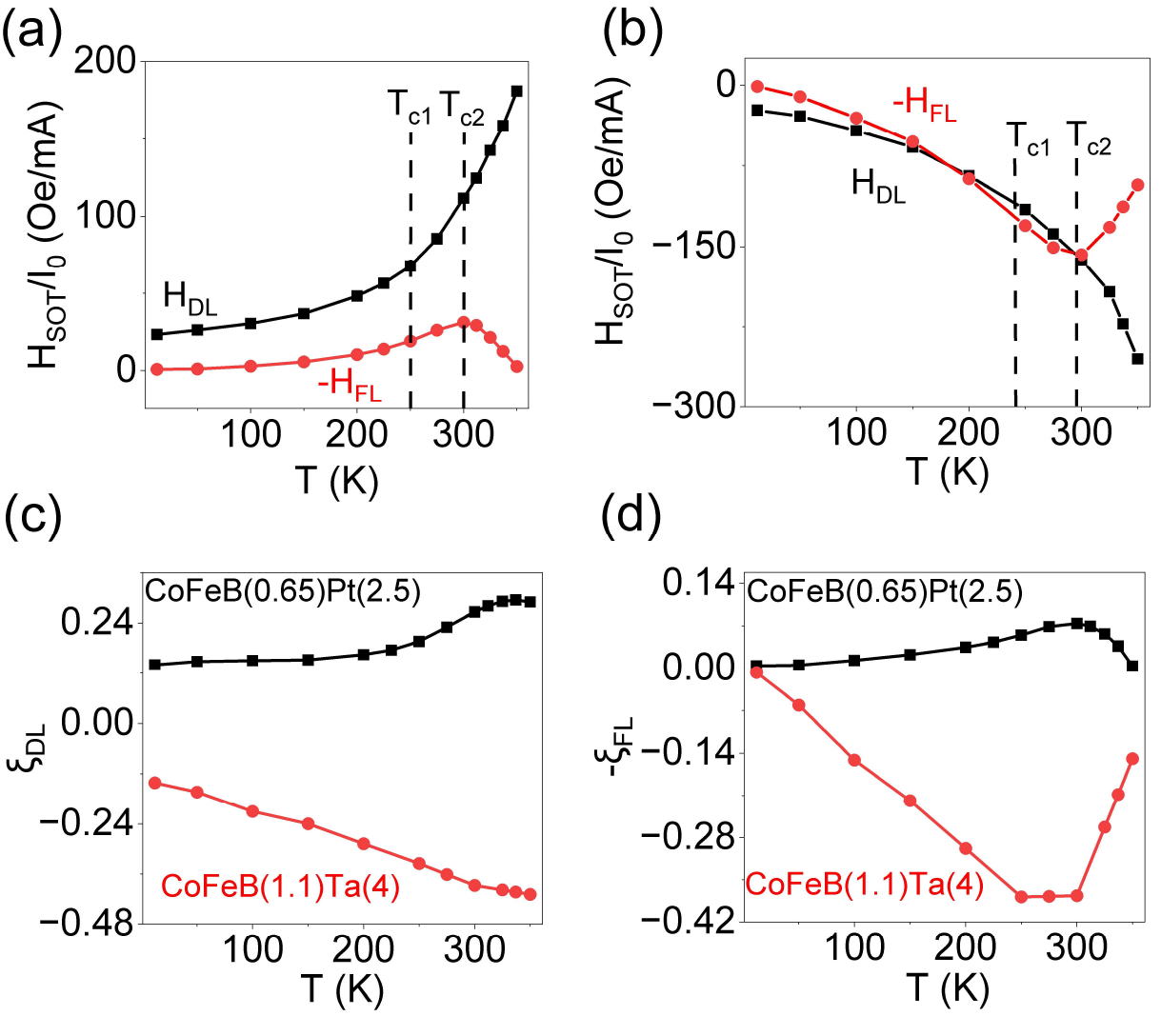}
		\caption{\label{fig:4}
Temperature dependence of the (a,b) damping-like and field-like effective fields and (c,d) SOT efficiencies for CoFeB(0.65)\slash Pt(2.5) (black) and CoFeB(1.1)\slash Ta(4) (red), extracted with the GMS method at $H=1$~kOe. Efficiencies $\xi_{\mathrm{DL,FL}}$ are computed from Eq.~(\ref{eq:efficiencies}) using $M(H,T)$ evaluated as described in Sec. \ref{sec:Characterization}. Dashed vertical lines mark $T_{C1}$ and $T_{C2}$.}
	\end{figure}

	The effect of fluctuations is elucidated by the temperature dependence of the SOFs at a fixed field, as shown in Fig.~\ref{fig:4}(a,b) for CoFeB(0.65)/Pt(2.5) and CoFeB(1.1)/Ta(4), respectively. The magnitudes of both $H_{\mathrm{DL}}$ and $H_{\mathrm{FL}}$ show a slow increase with increasing $T$ below the Curie temperature, consistent with the decreasing magnetization (see Fig.~\ref{fig:2}(c)), as described by Eq.~(\ref{eq:effective_fields}). The increase of $H_{\mathrm{DL}}$ accelerates above $T_{C1}$, while $H_{\mathrm{FL}}$ abruptly begins to decrease at $T>T_{C2}$, the critical temperature previously attributed to Rashba magnetism, as discussed in Section~\ref{sec:Characterization}.

	The observed dependences of $H_{\mathrm{DL}}$, $H_{\mathrm{FL}}$ on $T$ at $T>T_{C2}$ are consistent with the dependences on $H$ under the assumption that these dependences are determined by magnetization fluctuations. Since fluctuation amplitude decreases with increasing $H$ and increases with $T$, 
   an increase (decrease) of $H_{\mathrm{DL}}$ ($H_{\mathrm{FL}}$ ) with increasing fluctuation amplitude explains all the observed dependences.
	
	To further quantify the observed effects, we convert the effective fields into SOT efficiencies which represent the ratio of the spin current absorbed by the ferromagnet to the charge current density injected into the HM layer,~\cite{Manchon2019}
	\begin{equation}\label{eq:efficiencies}
		\xi_{\mathrm{DL,FL}}=\frac{2e\mu_0 M t_{\mathrm{FM}}H_{\mathrm{DL,FL}}}{\hbar j_c},
	\end{equation}
where $M$ is the field- and temperature-dependent magnetization evaluated as described in Sec.~\ref{sec:Characterization}. 

Figures~\ref{fig:4}(c,d) show the calculated $\xi_{\mathrm{DL,FL}}$ as a function of $T$. The calculated efficiencies  are within the broad range of values reported in the literature for Pt and Ta~\cite{Manchon2019}. In both samples, $\xi_{\mathrm{DL}}$ exhibits a gradual increase for $T<T_{C1}$. 
In this regime, the temperature dependence of $H_{DL}$ results mostly from the variation of magnetization. The DL efficiency exhibits a pronounced increase above $T_{C1}$, 
accounting for most of the increase of DL-SOF in the temperature range between $T_{C1}$ and $T_{C2}$. Meanwhile, the FL efficiencies increase from negligible values at deep cryogenic temperatures to a maximum at $T_{C2}$, and drop abruptly at higher $T$. These abrupt variations stand in stark contrast with  gradual changes of AHE hysteresis loops, confirming that the microscopic dynamical state of the magnetic system plays an active role in SOTs not limited to scaling of their effects by the magnetization.
	
\subsection{ST-FMR measurements}

	\begin{figure}[t]
		\includegraphics[width=\columnwidth]{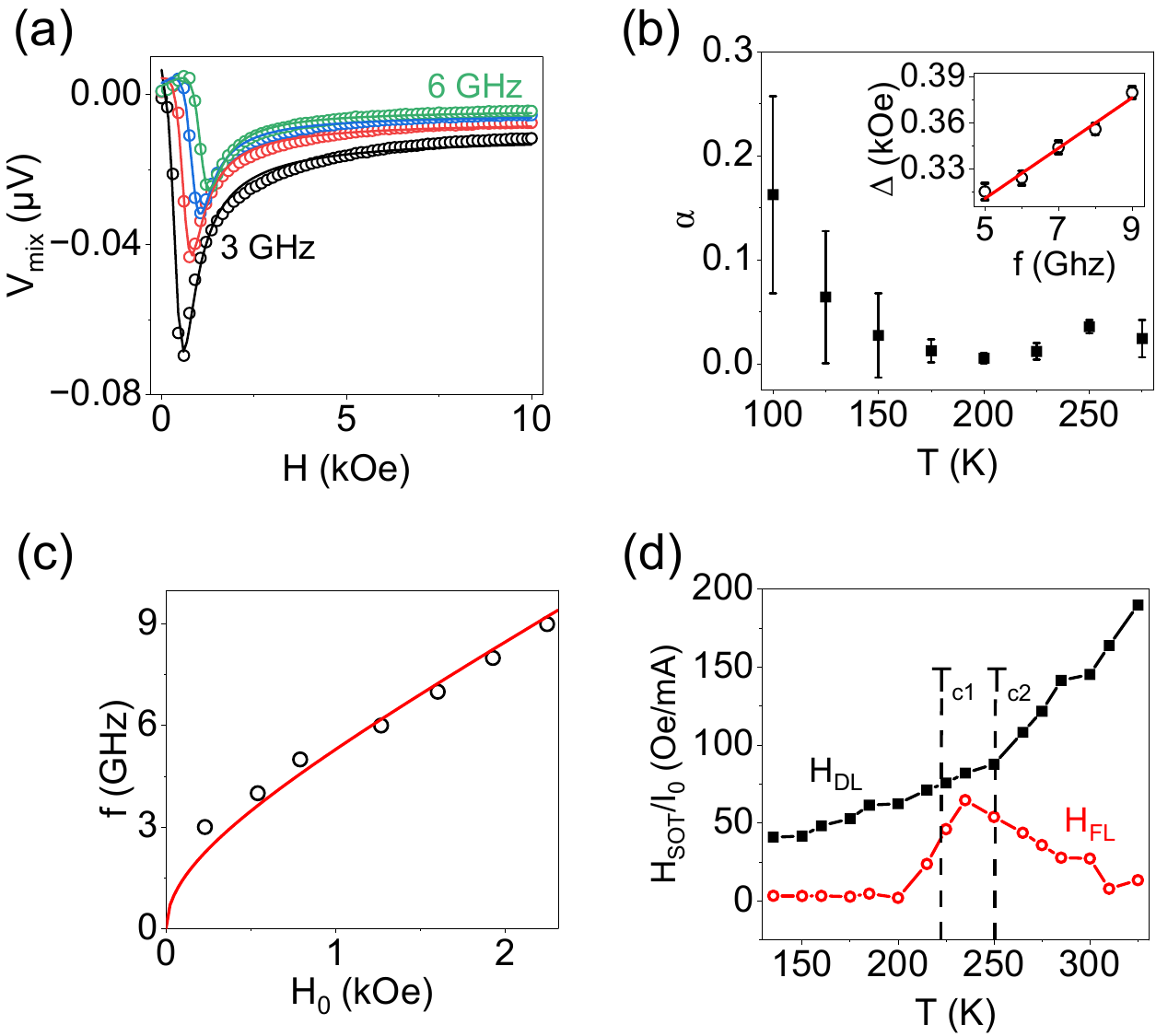}
		\caption{\label{fig:5}
			ST-FMR characterization of a Ta(1.5)/CoFeB(0.65)/Pt(2.5) Hall bar.
			(a) Mixing voltage $V_{\mathrm{mix}}$ vs $H$ at $T=275$~K for rf driving frequencies $f=3$--$6$~GHz in steps of $1$~GHz (symbols), fit to the sum of symmetric and antisymmetric Lorentzians (curves).
			(b) Gilbert damping $\alpha$ vs $T$ obtained from linear fits of the linewidth $\Delta$ vs $f$ (inset).
			(c) Resonance field $H_0$ vs $f$ fit to the  Kittel formula Eq.~(\ref{eq:kittel}), yielding the effective anisotropy field $H_a$.
			(d) Effective SOFs $H_{\mathrm{DL}}$ (closed symbols) and $H_{\mathrm{FL}}$ (open symbols) vs $T$ extracted as described in the text.}
	\end{figure}

As an independent test of the anomalous SOF behaviors in the vicinity of the Curie temperature, we performed spin-torque ferromagnetic resonance (ST-FMR) measurements in the Hall geometry~\cite{PhysRevApplied.14.024024}. In the ST-FMR method, an rf current $I_{\mathrm{rf}}\cos(2\pi f t)$ is applied to the Hall bar at frequencies $f$ close to the FMR frequency of the FM film, inducing SOT-driven magnetization precession. Mixing dc voltage $V_{\mathrm{mix}}$ produced by mixing between the resulting oscillation of the Hall resistance and rf current is recorded as a function of an in-plane field collinear with the current, and analyzed using a dynamical SOT model as described below.

We performed ST-FMR measurements on a Hall bar with the same width as the HHV samples and structure Ta(1.5)/CoFeB(0.65)/Pt(2.5), and the Curie temperature $T_{C1}=220$~K. The leads were patterned into microstrip lines enabling microwave-frequency measurements. The rf current amplitude $I_{\mathrm{rf}}=0.3$~mA was calibrated using known impedance mismatches and the line losses determined by microwave transmission measurements.

Before describing the ST-FMR results, we note that the interpretation of ST-FMR measurements for thermally fluctuating magnetization state is less straightforward than harmonic measurements, and does not permit quantitative analysis similar to low-frequency harmonic measurements discussed above. In particular, in the harmonic case the DL- and FL- effective fields drive in-plane and out-of-plane components of the magnetization that are distinguished through the distinct angular symmetries of their contributions to $V_{2\omega}(\varphi)$. By contrast, in ST-FMR both DL- and FL-SOTs drive the precession itself and cannot be uniquely distinguished based on the angular dependence. They can be separated through their different $t_{\mathrm{FM}}$ dependences, which is not practical for ultrathin films like ours. In the approach followed below, they are separated through their distinct contributions to the field dependence of the ST-FMR signal. However, near $T_{C}$ the fluctuating magnetization state also depends on the field, making this approach inherently less precise than HHV measurements due to the finite FMR width.

In the saturated state, $V_{\mathrm{mix}}(H)$ can be written as a sum of symmetric and antisymmetric Lorentzians~\cite{PhysRevApplied.14.024024},
	\begin{equation}\label{eq:ST-FMR}
		V_{\mathrm{mix}}(H)=V_s\,S(H)+V_a\,A(H),
	\end{equation}
	with $S(H)=\bigl[1+((H-H_0)/\Delta)^2\bigr]^{-1}$, $A(H)=(H-H_0)\Delta^{-1}S(H)$. Here $H_0$ is the resonance field, $\Delta=\alpha\omega/\gamma$ is the half-linewidth, $\alpha$ is the Gilbert damping, and $\gamma=1.76\times 10^{7}\,\mathrm{s^{-1}\,Oe^{-1}}$ is the gyromagnetic ratio. The amplitudes $V_{s,a}$ are related to the SOFs by~\cite{PhysRevApplied.14.024024}
	\begin{equation}\label{eq:Vmix}
		\begin{aligned}
			V_s&=\frac{I_{\mathrm{rf}}}{4\alpha H_+}\bigl(R_{\mathrm{AHE}}(H_{\mathrm{FL}}+H_{Oe})-R_{\mathrm{PHE}}H_{\mathrm{DL}}\bigr),\\
			V_a&=-\frac{I_{\mathrm{rf}}}{4\alpha H_+}\left[R_{\mathrm{PHE}}(H_{\mathrm{FL}} + H_{Oe})\frac{\gamma(H_0+H_a)}{\omega}+R_{\mathrm{AHE}}H_{\mathrm{DL}}\frac{\gamma H_0}{\omega}\right],
		\end{aligned}
	\end{equation}
	where $H_+=H_0+H_a/2$. Contributions from spin pumping and spin Seebeck/Nernst effects are negligible for CoFeB layers thinner than $\sim 6$~nm~\cite{PhysRevApplied.14.024024} and are not retained in our analysis.

	Because of fluctuation-induced field-dependent resonance broadening near $T_C$, $V_{\mathrm{mix}}(H)$ is expected to deviate from a pure Lorentzian. However, Eq.~(\ref{eq:ST-FMR}) empirically provides a good description of the data (Fig.~\ref{fig:5}(a)), allowing us to evaluate $H_0$, $V_{s,a}$, and $\Delta$.  We extract the effective Gilbert damping constant $\alpha$ from a linear fit of $\Delta$ vs $f$ (inset of Fig.~\ref{fig:5}(b)). It reaches a minimum $\alpha\sim0.005$ remarkably close to the value for bulk CoFeB at $T=200$~K somewhat below $T_C$ (Fig.~\ref{fig:5} (b)). Such a small damping value is surprising given that spin pumping is generally expected to enhance damping in  thin-film magnetic films~\cite{Nguyen2021,FMRGilbert}, which should be especially pronounced in the studied ultrathin films. The enhancement of damping near $T_C$ is expected due to the scattering on magnons whose populations diverge at the critical point, while its increase at low temperatures is likely due to the increased Pt-induced local magnetic anisotropy resulting in enhanced ST-FMR relaxation via two-magnon scattering. This interpretation is supported by the large fitting uncertainties at low $T$, as Lorentzian approximation of ST-FMR curves becomes increasingly inadequate.

	To extract SOFs from the ST-FMR amplitudes for the unsaturated state, we use the GMS method to express the prefactors in Eq.~(\ref{eq:Vmix}) in terms of the measured Hall susceptibilities,
	\begin{equation}\label{eq:amp_HSOT}
		R_{\mathrm{AHE}}\longrightarrow{H+H_a}\frac{dR_H}{dH_{\mathrm{OP}}},
		\qquad
		R_{\mathrm{PHE}}\longrightarrow\frac{H}{2}\frac{dR_H}{dH_{\mathrm{IP}}},
	\end{equation}
	both evaluated at $\theta_H=\varphi_H=0$, with $H_a$ obtained from a fit of $H_0(f)$ to the  Kittel formula~\cite{Kittel1948}
	\begin{equation}\label{eq:kittel}
		f=\frac{\gamma}{2\pi}\sqrt{H_0(H_0+H_a)}
	\end{equation}
	as shown in Fig.~\ref{fig:5}(c).

Substituting $V_{s,a}$ obtained at each $f,T$ into Eqs.~(\ref{eq:Vmix}) and~(\ref{eq:amp_HSOT}) yields the dependences of $H_{\mathrm{DL,FL}}$ on $T$ shown in Fig.~\ref{fig:5}(d). Despite the limitations of the ST-FMR method semi-empirically applied to non-saturated magnetization states, the temperature dependence of DL-SOF closely follows the results of HHV measurements. The FL contribution begins to increase at $T>200$~K, exhibits a peak close to the Curie point, and rapidly decreases at higher $T$, qualitatively reproducing the non-monotonic behaviors observed in the HHV measurements (Fig.~\ref{fig:4}(a,b)). However, in contrast to HHV results, the FL-SOF remains negligible at $T\leq 200$~K, and peaks between $T_{C1}$ and $T_{C2}$ instead of $T_{C2}$. Since precession dynamics intermixes DL and FL contributions, and their separation based on Eq.~(\ref{eq:Vmix}) in the ST-FMR sensitively depends on anisotropy, which does not have a well-defined value for the ultrathin films, the lack of quantitative agreement with the HHV method can be expected. Nevertheless, ST-FMR measurement independently confirms the divergence between the behaviors of DL- and FL-SOFs above the Curie temperature, with the former dramatically increasing and the latter becoming rapidly suppressed.

	%===============================================================================
	\section{Discussion}
	\label{sec:Discussion}
	%===============================================================================

Our central experimental finding is the dramatic divergence between $H_{\mathrm{DL}}$ and $H_{\mathrm{FL}}$ above the Curie temperature: $H_{\mathrm{DL}}$ is significantly enhanced, while $H_{\mathrm{FL}}$ is strongly suppressed. This finding is consistent with the expected effects of fluctuation-induced mixing between longitudinal spin conductance and transverse spin mixing conductance discussed in Section~\ref{sec:Theory}. Before further analyzing this mechanism, we consider possible alternative interpretations and additional contributions.

\textit{Scaling of SOFs with magnetization.} In the simplest interpretation, the strong dependence of SOFs on $T$ and $H$ near $T_C$ can be attributed to their inverse scaling with the temperature- and field-dependent magnetization. For fixed interface transparency, Eq.~(\ref{eq:effective_fields}) predicts that both $H_{\mathrm{DL}}$ and $H_{\mathrm{FL}}$ should scale as $1/M(H,T)$. However, analysis of SOT efficiencies that takes into account this contribution still shows a substantial increase of DL SOT efficiency above $T_{C1}$ (see Fig.~\ref{fig:4}(c)). Furthermore, a decrease of FL SOT efficiency at $T>T_{C2}$ is qualitatively inconsistent with this mechanism. Using $M(H,T)$ evaluated as described in Sec.~\ref{sec:Characterization}, we additionally computed the temperature dependence of SOFs expected from their scaling with magnetization (see Supplemental Material~\cite{supplemental}). The scaling accounts for the slow growth of $H_{\mathrm{DL}}$ at $T<T_C$, but does not reproduce its steep rise above $T_{C1}$. The calculated temperature dependence of $H_{FL}$ underestimates its increase observed at $T<T_{C1}$, suggesting a temperature-dependent electronic mechanism. The decrease of $H_{FL}$ observed at $T>T_{C2}$ is qualitatively inconsistent with the scaling.
We note that the variations of $\xi_{FL}$ above $T_C$ are too rapid and too large, and its behaviors are too similar between the two samples and distinct from $\xi_{DL}$ to be attributable to possible artifacts of our analysis such as the uncertainties in our evaluation of $M(H,T)$ or the limitations of GMS method used to extract the SOT efficiencies.

\textit{Averaging over fluctuating magnetization directions.} Equation~(\ref{eq:harmonic_2}) treats the magnetization as uniform and characterized by a single in-plane angle $\varphi$. However, the fluctuating magnetization effectively probes SOTs over a distribution of $\varphi$ determined by the temperature-dependent fluctuation amplitude. For broad angle distributions at large fluctuation amplitudes, such averaging preferentially suppresses higher harmonics, which could explain the decrease of $H_{FL}$. For $T>T_C$ the relation $M/M_s=\langle\cos\theta_f\rangle\approx 1-\langle\theta_f^2\rangle/2$ provides an estimate for the field dependence of the average fluctuation angle based on the field-dependent magnetization. If the fluctuations occur on a length scale larger than the spin diffusion length, the measured torque would be the average over the static configurations,
\begin{equation}\label{eq:av_torque}
	\langle\bm{\tau}\rangle=-\gamma\mu_0 M\,\langle\sin\theta_f\,\bm{H}_{\mathrm{SOT}}\rangle,
\end{equation}
which, for a dc field parallel to $\langle\bm{m}\rangle$, would predict a \emph{suppression} of SOT relative to the saturated case. However, the same averaging also reduces the response to the small transverse fields used to calibrate the SOFs. Thus, the effect of large-scale-fluctuations is already incorporated in the analysis, and cannot by itself explain the observed enhancement of DL-SOT or suppression of FL-SOT.

\textit{Temperature dependence of the spin Hall angle.} The spin Hall angle $\theta_{\mathrm{SH}}=\sigma_{\mathrm{SH}}\rho_N$ generally varies with $T$ due to the temperature dependences of both the spin Hall conductivity $\sigma_{SH}$ and of the resitivity $\rho_N$ of spin Hall material. However, in the dirty regime relevant to the studied ultrathin films, $\sigma_{\mathrm{SH}}$ is dominated by intrinsic and side-jump contributions that are weakly $T$-dependent~\cite{RevModPhys.82.1539}, and the temperature dependence of $\rho_N$ is too small (inset of Fig.~\ref{fig:2}(c)) to account for the order-of-magnitude change in $H_{\mathrm{DL}}$. The gradual temperature-dependent variations of SHE efficiency cannot explain the rapid variations of SOTs closely correlated with magnetic ordering. Furthermore, the close correspondence (aside from the flipped signs) between the temperature- and field-dependent behaviors for two distinct HMs (Pt and Ta) with opposite signs of spin Hall angles indicate that the observed effects cannot be explained by magnetoelectronic phenomena specific to these materials. In particular, strong temperature- and magnetization-dependence of proximity magnetism in Pt~\cite{Lim2013} could, in principle, result in large variations of SHE efficiency in Pt, but this would not explain similar variations in CoFeB/Ta where proximity magnetism is absent.

\textit{Rashba-Edelstein effect} is generally believed to contribute to SOTs, but the significance of this contribution in FM/HM bilayers has been debated~\cite{Tao2018,PhysRevLett.122.077201}. The sign reversal of $H_{\mathrm{FL}}$ at small $H$ in CoFeB/Pt, as well as CoFeB/Ta at cryogenic $T$, (insets in Figs.~\ref{fig:3}(a-c)) indicates an additional contribution besides the SHE-driven FL torque and too large for Oersted field, which is consistent with the presence of REE. Furthermore, an anomalous feature emerges in the AHE curves at $T>T_{C2}$, precisely the temperature at which FL-SOT begins to abruptly decrease. This feature was previously attributed to the competition between the exchange interaction and Bychkov-Rashba effective field  associated with the broken inversion symmetry at the FM/HM interface~\cite{Barnes2014,Kim2016,PhysRevApplied2024}. The close coincidence of the rapid $H_{\mathrm{FL}}$ drop with $T_{C2}$ points to the REE, which generates predominantly FL SOT~\cite{PhysRevB.89.220409}. According to the proposed mechanism of the competition between Rashba and exchange fields~\cite{PhysRevApplied2024}, the effective exchange field is dominant below $T_{C2}$ so that the SOT is well described by the SHE alone~\cite{PhysRevApplied.14.024024}. At $T>T_{C2}$, the effective Rashba field becomes large, resulting in a rapidly varying contribution to FL-SOT. 

In principle, this mechanism can provide a natural explanation for the rapid suppression of $H_{\mathrm{FL}}$ at $T>T_{C2}$ and its sign reversal at small $H$. However, it requires a precise cancellation of the SHE contribution by the REE contribution, for two different material combinations with very different interface properties. In CoFeB/Pt, the interface is proximity magnetized, while CoFeB/Ta exhibits a magnetically dead layer. Therefore, the effects of temperature-dependent interplay between magnetism and Rashba SOC for these different material combinations should be different. We also note that in all measured samples $H_{\mathrm{FL}}\to 0$ as $T\to 0$, while DL-SOT remains finite. Since a finite FL-SOT is generally expected for the SHE mechanism, it is possible that in the low-$T$ limit, the REE contribution to FL-SOT nearly cancels the SHE contribution. However, this again requires precise cancellation between the magnitudes of two distinct effects, for two distinct material combinations, which happens precisely in the $T\to0$ limit. Such a fortuitous coincidence is highly unlikely, leading us to conclude that while REE may contribute to the observed SOT, it is unlikely to be the dominant origin of the observed anomalous behaviors in the vicinity of $T_C$. 

%In the semi-classical models of SHE-driven SOTs, FL-SOT is generally concomitant with DL-SOT as determined by the dynamics of non-collinear conduction-electron spin component~\cite{Manchon2019}, so their relation is not expected to dramatically change with $T$. It is possible that the SOT picture based on the semi-classical s-d exchange model may become inadequate for ultrathin films at deep cryogenic temperatures, perhaps due to the dominance of  non-classical spin transfer effects~\cite{Zholud2017} or many-body correlations~\cite{Ivanov2023}. Its mechanisms warrant further investigations beyond the scope of this work. 

\begin{figure}[t]
\includegraphics[width=0.6\columnwidth]{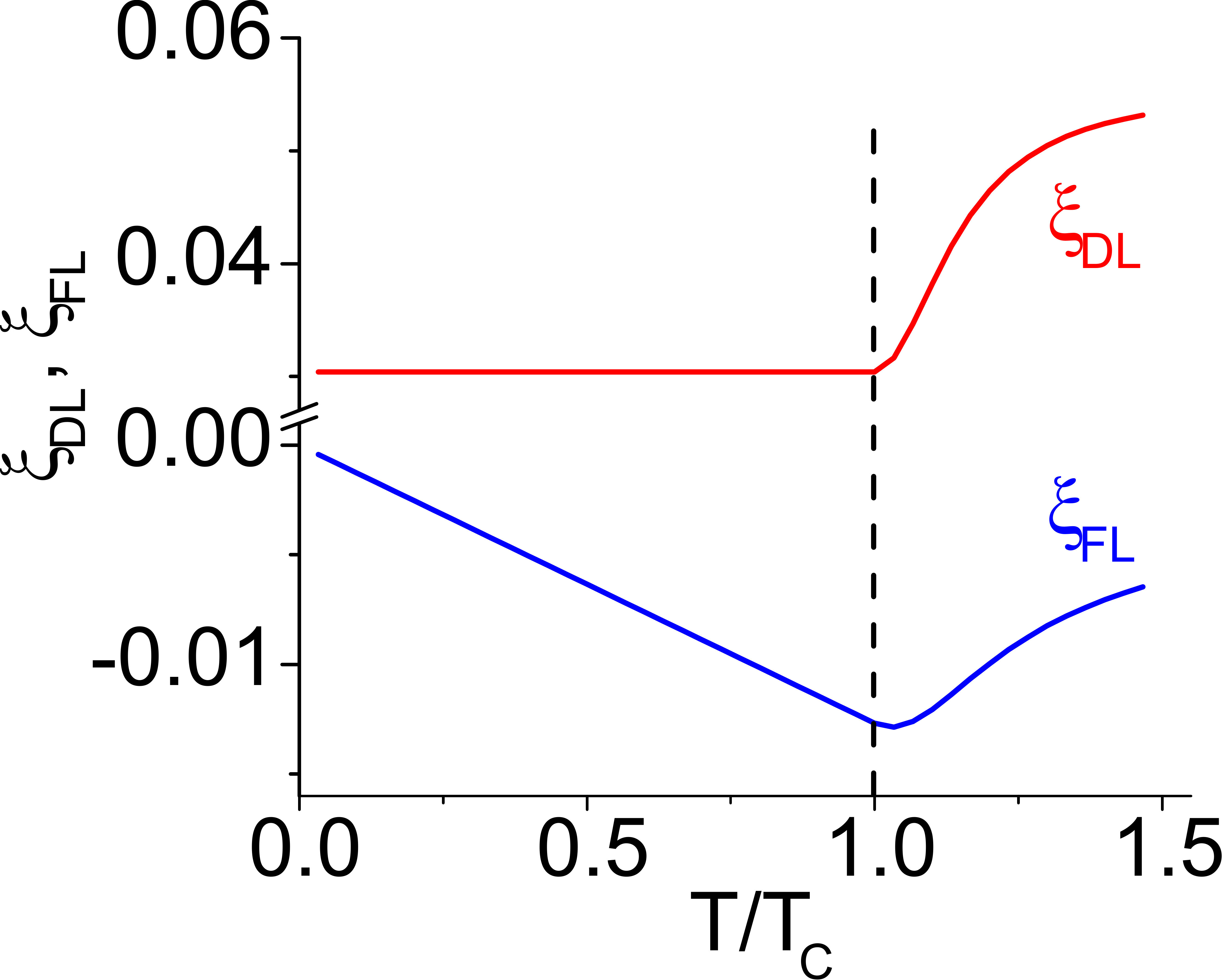}
\caption{\label{fig:simulation}
DL and FL SOT efficiencies vs $T/T_C$ calculated based on Eq.~(\ref{eq:cond_tensor_av}) with negligible $G_r$, $G_i=aT$ with constant $a$ and $D$ in Eq.~(\ref{eq:cond_tensor_av}) adjusted to match the experimental data scale for CoFeB/Pt. The averages over fluctuation angle $\theta_f$ were calculated using Boltzmann distribution for the magnetic moment $m=k_BT/\mu_0\Delta H$, with $\Delta H(T)$ determined as described in Section~\ref{sec:Characterization}.}
\end{figure}

\textit{Fluctuation-induced spin-conductance mixing.} The short-length-scale fluctuation picture of Sec.~\ref{sec:Theory}, encoded in Eq.~(\ref{eq:cond_tensor_av}), predicts an enhancement of the real (DL-related) part of the spin-mixing conductance and a suppression of the imaginary (FL-related) part, precisely as observed in our measurements. The relative enhancement of $\xi_{\mathrm{DL}}$ at $T>T_{C}$ is noticeably stronger in CoFeB/Pt, consistent with the presence of fluctuating proximity-induced magnetic moments at the Pt interface that can participate in the fluctuation-driven mixing, in contrast to the magnetically dead layer at the Ta interface. 

To further validate this mechanism, we performed semi-phenomenological calculations of the temperature-dependence of SOT efficiencies based on the short-lengthscale fluctuation model of Section~\ref{sec:Theory}. We utilize Langevin-Brillouin model of independent fluctuating nanoscale magnetic moments. The Boltzmann distribution gives the fluctuation angle distribution $P(\theta_f)=\frac{x}{2\sinh x}e^{x\cos\theta_f}\sin\theta_f$, where $x=m\mu_0H/k_BT$, $k_B$ is the Boltzmann constant, and the fluctuating magnetic moment $m=k_BT/\mu_0\Delta H$ is determined from the temperature-dependent broadening $\Delta H$ of the hysteresis curves discussed in Section~\ref{sec:Characterization}.

 The fluctuation-induced scaling factor in the imaginary spin mixing component of the averaged spin conductance tensor is $\langle\cos\theta_f \rangle=L(x)$, where $L(x)=\coth x-1/x$ is the Langevin function. The corresponding fluctuation-induced scaling factor in the real component of spin mixing conductance is $\langle\sin^2\theta_f \rangle=2L(x)/x$. Figure~\ref{fig:simulation} shows the results of calculations of SOTs based on $\Delta H(T)$ approximated by Eq.~(\ref{eq:width}) with $\Delta H_{in}=1.2$~kOe, neglected anisotropy ($H_a=0$), $a=40$~Oe/K, the overall scale of efficiencies adjusted to match the observations, with neglected $G_r$, and $G_i$ taken to be proportional to $T$ to match the observed linear temperature dependence of FL-SOT. These calculations reproduce the overall anomalous behaviors of SOT efficiencies in CoFeB/Pt shown in Figs.~\ref{fig:4}(c),(d), including a kink-like enhancement of DL-SOT and the rapid suppression of FL-SOT. However, in our model both effects appear at $T>T_{C1}$, while in our  measurements the drop of FL-SOT is observed only above $T_{C2}$, and is more rapid than expected from the model. 
 
 We note that all the analysis discussed above is based on the spin models of magnetism and spin-orbitronic phenomena. While the importance of orbital degrees of freedom in single-particle phenomena is becoming increasingly recognized, their role in collective phenomena such as magnetic ordering and fluctuations is only starting to emerge~\cite{Ivanov2023}. We speculate that some of the effects observed in this work may be influenced by the interplay between collective orbital phenomena, magnetism and spin transport, warrant further studies of magnetic and magnetoelectronic phenomena near the critical temperatures. 

	%===============================================================================
	\section{Summary and outlook}
	\label{sec:Summary}
	%===============================================================================

	In this work, we have studied SOTs in ultrathin ferromagnetic films whose Curie temperature is suppressed due to confinement. To enable the analysis of the fluctuating-magnetization regime above the Curie point, we have developed a self-consistent generalized magnetoelectronic susceptibility framework. Unlike standard Hall harmonic voltage analysis, the GMS method bypasses the assumptions of magnetic saturation and uniform anisotropy, making it a robust metrological tool for ultrathin films operating near $T_C$. The same approach is extended to ST-FMR; although the latter is inherently less precise than harmonic analysis in the fluctuating regime, it provides an independent corroboration of the GMS results.

	Our central experimental finding is the divergent temperature dependence of the two SOFs in CoFeB/Pt and CoFeB/Ta near and above $T_C$: $H_{\mathrm{DL}}$ rises sharply, while $H_{\mathrm{FL}}$ is suppressed and can reverse sign. The enhancement of $H_{\mathrm{DL}}$ is consistent with a fluctuation-driven mixing of the longitudinal spin conductance into the transverse channel at the FM/HM interface. The suppression of $H_{\mathrm{FL}}$ is also consistent with the fluctuation-driven mixing between the components of spin conductance tensor, but the observed effects are larger than expected, suggesting the presence of additional mechanism such as the effective Rashba-Edelstein field that emerges once the exchange interaction is sufficiently weakened by proximity to the magnetic transition, or anomalous interplay between orbital and spin states.

	From an applications standpoint, the high current densities required for SOT-driven magnetization switching unavoidably heat the magnetic layer, in addition to the direct excitation of high-frequency magnons by SOT that can be approximated as effective magnetic temperature increase~\cite{Demidov2011,Demidov2017}. Our results suggest that, rather than mitigating this Joule heating and high-frequency magnon excitation, one can exploit it: by engineering heterostructures with tailored $T_C$ modestly above the device operation temperature, one could reach the fluctuation-enhanced regime analogous to heat-assisted magnetic recording. 
    This mechanism requires careful balancing with the thermal device stability.  In addition to enhancement of spin-mixing conductance, enhanced fluctuations also reduce the net magnetization, resulting in a trivial but useful enhancement of effective spin-orbit fields. More broadly, since the enhancement of $\xi_{\mathrm{DL}}$ originates from the geometric averaging of the interfacial spin conductance over a fluctuating microscopic magnetic state, the same effect should be accessible below $T_C$ through static, engineered inhomogeneity --- e.g., via frustrated exchange coupling, controlled interfacial disorder, or coupling to an antiferromagnetic spacer.
	
	\begin{acknowledgments}
		We thank Martin Mourigal for providing assistance with SQUID magnetometry. The contributions of M.S. and S.U. were supported by the NSF award ECCS-2448290, and S.I. by the Emory URC award.
	\end{acknowledgments}

	\section*{Data availability}
	The data supporting the findings of this study are available at Zenodo (10.5281/zenodo.21114381).

\begin{comment}

\end{comment}

	\bibliography{references}
\end{document}